\documentclass[aps,prb,showpacs,twocolumn]{revtex4}
\usepackage{graphicx}
\usepackage{bm}
\begin{document}
\title{Direct-current control of radiation-induced differential magnetoresistance oscillations
in two-dimensional electron systems}
\author{X. L. Lei}
\affiliation{Department of Physics, Shanghai Jiaotong University,
1954 Huashan Road, Shanghai 200030, China}

\begin{abstract}
Magnetoresistance oscillations in two-dimensional electron systems driven 
simultaneously by a strong direct current and a microwave irradiation,
are analyzed within a unified microscopic scheme treating both 
excitations on an equal footing.
The microwave-induced resistance oscillations are described by
a parameter $\epsilon_\omega$ proportional to the radiation frequency,
while the dc-induced resistance oscillations are  
governed by a parameter $\epsilon_j$ proportional to the current density.
In the presence of both a microwave radiation and a strong dc,
the combined parameter $\epsilon_\omega+\epsilon_j$ is shown to control 
the main resistance oscillations, in agreement with the recent 
measurement [Zhang {\it et al.} Phys. Rev. Lett. {\bf 98}, 106804 (2007)].

\end{abstract}

\pacs{73.50.Jt, 73.40.-c, 73.43.Qt, 71.70.Di}

\maketitle

Interest in the direct current effect on magnetoresistance in two-dimensional (2D)
electron system (ES) has recently been revived due to the exciting experimental observations   
that in the case without irradiation, a relatively weak current can 
induce substantial magnetoresistance oscillations and negative differential 
resistivity.\cite{Yang2002,Mani2004,Bykov,JZ-prb07}
Recent careful measurements\cite{WZ-prb07} and theoretical analysis \cite{Lei07}
disclosed accurate oscillating charateristic having the period $\Delta\epsilon_j\sim 1$, expressed 
by a dimensionless parameter $\epsilon_j$ proportional to the ratio of the 
current density to the strength of the magnetic field. 

The main oscillations appearing in the
linear microwave photoresistance were known to be periodic in 
$\epsilon_\omega=\omega/\omega_c$ ($\omega$ is the angular frequency of
the radiation and $\omega_c=eB/m$ is the cyclotron frequency of an effective-mass $m$ electron
in the magnetic field $B$), having the period 
$\Delta\epsilon_\omega=1$.\cite{Ryz,Zud01,Ye,Mani,Zud03,Dor03,Willett,Durst,Lei03,
Dmitriev03,Ryz03,Vav,Lei05,Lei06}  Recent studies by Zhang {\it et al.}\cite{WZ-prl07}
 on magneto-photoresistance in a 2DES subject to 
a microwave irradiation and a finite direct-current excitation revealed 
 that the maxima (minima) of the differential resistance  
can evolve into minima (maxima) and back, and the oscillations are 
governed by a parameter $\epsilon= \epsilon_\omega +\epsilon_j$, combining the  
dc related parameter $\epsilon_j$ and the radiation-frequency related 
parameter $\epsilon_\omega$.
 
The preliminary explanation of the observed resistance oscillations 
under simultaneous dc and ac excitations, which is based on 
indirect electron transitions consisting of a radiation-induced 
jump in energy and a dc-induced jump in space between Hall-field-tilted 
Landau levels,\cite{WZ-prl07} is obviously not satisfied.
A systematic theory capable of treating both excitations simultaneously
within a single framework is highly desirable.

It is better to treat the magnetotransport subject to a strong dc excitation 
with a scheme direct using the electric current, rather than field, 
as the basic control parameter. Such a theory has been developed.\cite{Lei03,Lei05}
In this theory the integrative drift motion of the electron system 
opens new channels for individual electron to transit 
between different Landau levels through impurity and phonon scatterings, 
thus the direct current and the microwave radiation are dealt with
on an equal footing.

We consider a 2DES having $N_{\rm s}$ electrons in a unit area of 
the $x$-$y$ plane subject to a uniform magnetic field ${\bm B}=(0,0,B)$ 
in the $z$ direction.
When an electromagnetic wave with incident electric field ${\bm E}_{{\rm i}s}\sin \omega t$,
irradiates the 2D plane together with a dc electric field ${\bm E}_0$ inside,
the steady transport state of the system can be described by the electron drift velocity
${\bm v}_0$ and an electron temperature $T_{\rm e}$, satisfying the 
force and energy balance equations:\cite{Lei05}
\begin{eqnarray}
N_{\rm s}e{\bm E}_{0}+N_{\rm s} e ({\bm v}_0 \times {\bm B})+
{\bm F}_0&=&0,\label{eqv0}\\
N_{\rm s}e{\bm E}_0\cdot {\bm v}_0+S_{\rm p}- W&=&0.
\label{eqsw}
\end{eqnarray}
Here, the frictional force  resisting electron drift motion,
\begin{equation}
{\bm F}_0=\sum_{{\bm q}_\|}\left| U({\bm q}_\|)\right| ^{2}
\sum_{n=-\infty }^{\infty }{\bm q}_\|{J}_{n}^{2}(\xi ){\it \Pi}_{2}
({\bm q}_\|,\omega_0+n\omega ),\label{ff0}
 \label{eqf0}
\end{equation}
is given by the electron density correlation function ${\it \Pi}_2({\bm q}_{\|},{\it \Omega})$,
the effective impurity potential $U({\bm q}_{\|})$, a radiation-related coupling parameter
$\xi$ in the Bessel function $J_n(\xi)$, and $\omega_0\equiv{\bm q}_{\|}\cdot {\bm v}_0$.
The electron energy absorption from the radiation field, $S_{\rm p}$, and the electron energy dissipation 
to the lattice, $W$, are given in Ref.\,\onlinecite{Lei05}.
The nonlinear longitudinal resistivity and differential resistivity 
in the presence of a radiation field are direct obtained from Eq.\,(\ref{eqv0}) 
by taking ${\bm v}_0$ (i.e the current ${\bm J}=N_{\rm s}e{\bm v}_0 $) 
in the $x$ direction, ${\bm v}_0=(v_{0},0,0)$ and ${\bm J}=(J,0,0)$:
\begin{eqnarray}
R_{xx}&=&-{F}_0/(N_{\rm s}^2 e^2 v_{0}),\label{rxx}\\
r_{xx}&=&=-({\partial F_0}/{\partial v_0})/(N_{\rm s}^2 e^2).\label{rrxx}
\end{eqnarray}

In the absence of microwave radiation, the formulation reduces to those
presented in Ref.\,\onlinecite{Lei07}. The initial suppression and subsequent 
oscillation of the magnetoresistivity arise from the current-assisted electron 
transitions and are controlled by the parameter
\begin{equation}
\epsilon_j\equiv \frac{\omega_j}{\omega_c}=\frac{2mk_{\rm F}v_0}{eB}
=\sqrt{\frac{8\pi}{N_{\rm s}}}\frac{m}{e^2}\frac{J}{B}.
\end{equation}
The oscillating valley-peak pairs in the $R_{xx}$-vs-$\epsilon_j$ curve
appear around the positions $\epsilon_j=\eta m\,\,\,(m=1,2,...)$, with $\eta\sim 1$, depending 
on the form of the scattering potential and on the electron temperature.

In the case of linear photoresistance, ${\bm v}_0\rightarrow 0$, the electron transition
(intra- and inter-Landau levels) can take place 
 by absorbing or emitting $n$ photons of frequency $\omega$
 and jumping from Landau level $l$ to $l'$ (across $m$ levels). 
 This process gives rise to terms of $\pm n$
 in the frictional force ${\bm F}_0$ expression Eq.\,(\ref{eqf0}).
 The periodicity of the electron density correlation function  
${\it \Pi}_2({\bm q}_\|, {\it \Omega}+\omega_c)={\it \Pi}_2({\bm q}_\|, {\it \Omega})$
at low electron temperature and many Landau-level occupation, 
leads to the appearance of peak-valley pairs in
the linear magnetoresistivity around the positions of $n\omega=m\omega_c$.\cite{Lei06}
The primary peak-valley pairs come from the single-photon process, $n=1$,
showing up around the node positions $\omega=m\omega_c$, 
or 
\begin{equation}
\epsilon_{\omega}\equiv{\omega}/{\omega_c}=m=1,2,3,\cdots, 
\end{equation}
and the primary period of $R_{xx}$ oscillation is 
$\Delta\epsilon_\omega=1$. Two- and multiple-photon
processes ($n\geq 2$) yield secondary peak-valley pairs.\cite{Lei06}

When there is a finite bias current in 2DES, in view of 
the extra energy $\omega_0\equiv {\bm q}_\|\cdot{\bm v_0}$
provided by the electron drift motion together with the energy $n\omega$ 
provided by the absorption or emission of $n$ photons of the radiation field, 
an electron can be scattered by elastic impurities and jumps across $m$ Landau levels
of different energies. The resonance condition in this case should  be
determined by $\eta \,\omega_j+n\omega=m\omega_c$. Focusing only on 
single-photon processes ($n=1$), the valley-peak pairs of  
nonlinear magnetoresistance oscillations are anticipated to appear around the positions
\begin{equation}
\epsilon_\omega+\eta\, \epsilon_j=\epsilon_\omega \left(1+\eta{\omega_j}/{\omega}\right)=m
\end{equation}
or
\begin{equation}
\epsilon_\omega\equiv{\omega}/{\omega_c}={m}/({1+\gamma_j}),
\end{equation}
where $m=1,2,3,\cdots$, and 
\begin{equation}
\gamma_j\equiv \eta \,{\omega_j}/{\omega}.
\end{equation}
Therefore, the oscillating behavior of $R_{xx}$ 
is controlled by the parameter $\epsilon_\omega+\eta \,\epsilon_j$,
or $\epsilon_\omega (1+\gamma_j)$. 

\begin{figure}
\includegraphics [width=0.48\textwidth,clip=on] {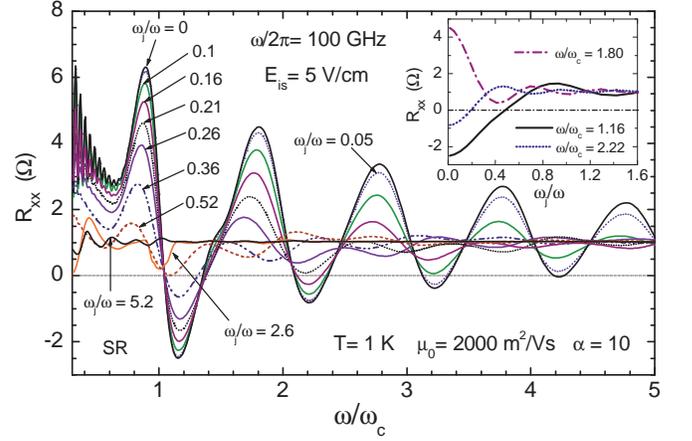}
\vspace*{-0.2cm}
\caption{(Color online) Longitudinal magnetoresistivity $R_{xx}$
vs $\epsilon_{\omega}=\omega/\omega_{c}$ for a GaAs-based 2DES 
with  $N_{\rm s}=3\times 10^{15}$/m$^{2}$,
 $\mu_0=2000$\,m$^{2}$/V\,s and $\alpha=10$
at lattice temperature $T=1$\,K,
under linearly polarized irradiation of
frequency $\omega/2\pi=100$\,GHz and incident amplitude $E_{{\rm i}s}=5$\,V/cm
and subject to bias current densities of 
$\omega_j/\omega=0, 0.05, 0.1, 0.16, 0.21, 0.26, 0.36, 0.52, 2.6$, and 5.2.
The elastic scatterings are due to short-range impurities.
The inset shows $R_{xx}$ as a function of $\omega_j/\omega$ 
at three fixed magnetic field strengths,
$\omega/\omega_c=1.16,1.80$, and 2.22.}
\label{fig1}
\end{figure}

Figure 1 presents the calculated longitudinal magnetoresistivity $R_{xx}$
vs $\epsilon_{\omega}=\omega/\omega_{c}$ for a GaAs-based 2DES
with carrier density $N_{\rm s}=3\times 10^{15}$/m$^{2}$, low-temperature
linear mobility $\mu_0=2000$\,m$^{2}$/V\,s 
at lattice temperature $T=1$\,K,
irradiated by a linearly $x$-direction polarized microwave of
frequency $\omega/2\pi=100$\,GHz and incident amplitude $E_{{\rm i}s}=5$\,V/cm,
 and subject to bias velocities 
$2v_0/v_{\rm F}=0, 0.001, 0.002, 0.003, 0.004, 0.005, 0.007, 0.01, 0.05$, and 0.1 
($v_{\rm F}$ is the Fermi velocity), corresponding to 
current densities $J=0.06, 0.11, 0.17, 0.23, 0.29, 0.40, 0.57, 2.9$, and 5.7 A/m
or the bias parameter 
$\omega_j/\omega=0, 0.05, 0.1, 0.16, 0.21, 0.26, 0.36, 0.52, 2.6$, and 5.2, respectively.
The elastic scatterings are assumed due to short-range impurities and 
the Landau level broadening parameter is taken to be $\alpha=10$.\cite{Lei05}
Linear (zero dc bias, $\omega_j=0$) magnetoresistivity exhibits typical feature of 
radiation-induced magnetoresistance oscillations, 
with two-photon process slightly showing up as a shoulder around 
$\epsilon_\omega\approx 1.5$.   
In the case of small dc bias, $\gamma_j\ll 1$, the node positions of peak-valley 
pairs are shifted down from  $\epsilon_\omega=m$ at zero dc bias 
to $\epsilon_\omega=m/(1+\gamma_j)\approx m-m\gamma_j$. The shifted distances 
relative to the original positions, $m\gamma_j$, are larger for larger $m$, but
still locate within the same order of $\epsilon_\omega$  
as long as $m\gamma_j < 1$. When the shifted distance becomes as large as $0.5$, 
the original peak will move to the range where previously there is a valley.
In the case of strong dc bias, $\gamma_j > 1$, the original pairs of higher order
(larger $m$) may shift to lower order range of $\epsilon_\omega$.
For instance, in the case of short-range impurity scattering $\eta=0.94$,\cite{Lei07}
for the bias current of $\omega_j/\omega=5.2$ ($\gamma_j=4.9$), the original pairs at
$\epsilon_\omega=m$ with $m=1,2,3,4,5$, and 6 are shifted to the positions 
$\epsilon_\omega=m/(1+\gamma_j)=0.17,0.34,0.51,0.68,0.85$, and 1.02.
Except for the first one ($\epsilon_\omega=0.17$), which is beyond the range plotted, the other
five pairs are clearly identified in the figure. 

The inset of Fig.\,1 shows the change of the magnetoresistivity $R_{xx}$ with increasing bias
current density $J$ in terms of $\omega_j/\omega$ at three fixed magnetic field strengths
$\omega/\omega_c=1.16,1.80$, and 2.22, respectively, near the first minimum, the second 
maximum and minimum of $R_{xx}$. Here we show the fact that 
within a certain magnetic field range of resistivity minimum,  $R_{xx}$ 
can be negative at small $J$ but increases with increasing $J$ 
and passes through zero at a finite $J$.\cite{Lei03} This is exactly what
is required for the instability of the time-independent small-current solution 
and for the development of a spatially nonuniform\cite{Andreev,Alicea05} or
a time-dependent state,\cite{Ng} which exhibits measured zero resistance.
Therefore, the region where an absolute negative dissipative magnetoresistance 
develops is identified as that of zero-resistance state.

\begin{figure}
\includegraphics [width=0.46\textwidth,clip=on] {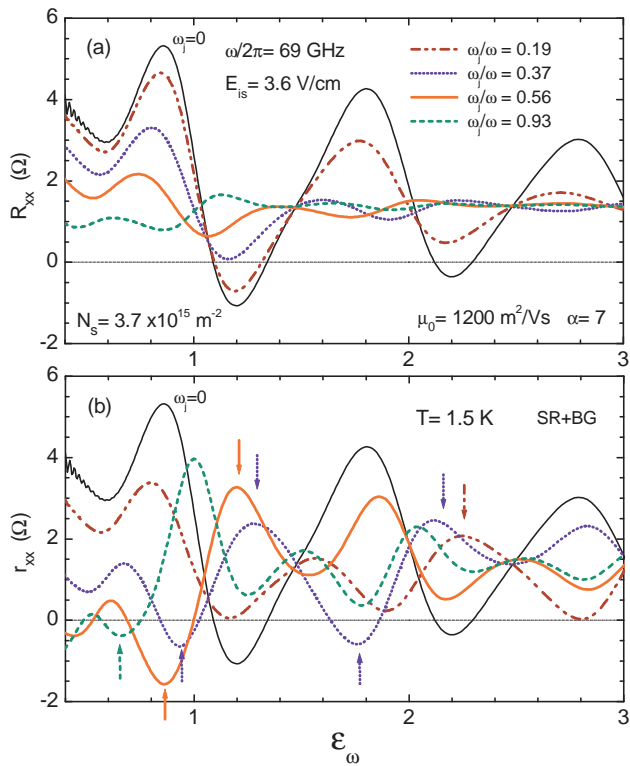}
\vspace*{-0.2cm}
\caption{(Color online) Longitudinal magnetoresistivity $R_{xx}$ (a) and 
differential magnetoresistivity $r_{xx}$ (b) vs  
$\epsilon_{\omega}\equiv \omega/\omega_{c}$ for a GaAs-based 2DES 
with $N_{\rm s}=3.7\times 10^{15}$/m$^{2}$ and $\mu_0=1200$\,m$^{2}$/V\,s 
at $T=1.5$\,K, under irradiation 
of a linearly polarized microwave of frequency $\omega/2\pi=69$\,GHz 
and incident amplitude $E_{{\rm i}s}=3.6$\,V/cm, and 
subject to bias current densities of 
$\omega_j/\omega=0,  0.19,  0.37,  0.56$, and 0.93. The elastic scatterings 
are from a mixture of short-range (SR) and background (BG) impurities.}
\label{fig2}
\end{figure}

The calculated longitudinal magnetoresistivity $R_{xx}$ and 
differential magnetoresistivity $r_{xx}$ for another GaAs-based 2DES 
with carrier density $N_{\rm s}=3.7\times 10^{15}$/m$^{2}$ and low-temperature
linear mobility $\mu_0=1200$\,m$^{2}$/V\,s 
at lattice temperature $T=1.5$\,K are plotted in Fig.\,2 as functions of 
$\epsilon_{\omega}\equiv \omega/\omega_{c}$, under the irradiation 
of a linearly $x$-direction polarized microwave of frequency $\omega/2\pi=69$\,GHz 
with incident amplitude $E_{{\rm i}s}=3.6$\,V/cm, and 
subjected to bias drift velocities  
$2v_0/v_{\rm F}=0,  .002,  .004,  .006$, and 0.01, 
corresponding to current densities $J=0, 0.14, 0.28, 0.42$, and 0.70 A/m or 
 bias parameter
$\omega_j/\omega=0,  0.19,  0.37,  0.56$, and 0.93, respectively. 
The elastic scatterings are assumed due to a mixture of short-range and background
impurities and the broadening parameter is taken to be $\alpha=7$.
We see that positions of peak-valley pairs and their maxima and minima 
of different $\epsilon_{\omega}$ orders $m$ in the zero-bias $R_{xx}$ curves [Fig.\,2(a)] 
are shifted downward with increasing current density 
$\omega_j/\omega$ following the same rule as stated in the case of Fig.\,1.

The oscillations of differential resistivity $r_{xx}$
exhibit shorter period and much larger amplitude than those of $R_{xx}$,
especially at high dc bias where $R_{xx}$ oscillations become less prominent, as shown in Fig.\,2(b). 
Positions of peak-valley pairs and their maxima and minima in  
$r_{xx}$  are shifted downward further relative to
those of the corresponding $R_{xx}$ at the same order $m$.
Of particular interest is that 
with increasing bias current density, in the $\epsilon_\omega$ range where
 the zero-bias magnetoresistivity ($R_{xx}=r_{xx}$) exhibits positive maximum,  
the differential magnetoresistivity $r_{xx}$ can be driven down to 
a considerable negative value, as indicated by the up-directed arrows
in Fig.\,2(b); and  in the $\epsilon_\omega$ range where
 the zero-bias magnetoresistivity ($R_{xx}=r_{xx}$) exhibits a minimum,   
the differential magnetoresistivity $r_{xx}$ can be driven up to 
exhibit a maximum, as indicated by the down-directed arrows
in Fig.\,2(b). 

According to electrodynamic analyses, a
homogeneous state of current $J$ is unstable if either the absolute
dissipative resistivity $R_{xx}$ or the differential resistivity $r_{xx}$
becomes negative.\cite{Andreev,Alicea05} Thus, despite the detailed structure yet to be explored,
it is expected that, under appropriate conditions,
 a zero-resistance state (ZRS) 
or some kind of quasi-ZRS could be measured in the region where $R_{xx}$ or $r_{xx}$ 
exhibits a negative value in a homogeneous microscopic analysis. 
With this in mind, the $R_{xx}$ and $r_{xx}$
behaviors in Fig.\,2 could predict the following measured results.
In the vicinity around $\epsilon_\omega=2.25$, the ZRS showing up in the zero dc bias
will disappear at bias $\omega_j/\omega \geq 0.10$ because both $R_{xx}$ and $r_{xx}$ 
return to positive. Instead, one should observe a bulged $r_{xx}$ in the cases of 
$\omega_j/\omega = 0.19, 0.37$, and 0.93  and a dented $r_{xx}$ 
in the case of $\omega_j/\omega = 0.56$, as shown in Fig.\,2.
In the vicinity around $\epsilon_\omega=1.25$, where the negative $R_{xx}$ (thus the ZRS)
maintains up to the dc bias $\omega_j/\omega = 0.35$, one may not detect $r_{xx}$ 
significantly different from zero at low dc biases but the bulged $r_{xx}$ should 
be observed at higher dc biases when $R_{xx}$ becomes positive and the ZRS disappears,
as seen in Fig.\,2 in the cases of $\omega_j/\omega = 0.37$ and 0.56.
In the vicinity around $\epsilon_\omega=1.75$ and 0.8, the zero-bias  
magnetoresistivity $R_{xx}=r_{xx}$ exhibits main peaks  
while the differential resistivity $r_{xx}$ becomes negative, respectively, at 
finite dc biases $\omega_j/\omega = 0.37$ and 0.56,
suggesting the possible ZRS induced by the direct current.  
When further increasing current density, 
it disappears and becomes a bulged differential resistivity.
In the range $0.6\leq\epsilon_\omega\leq 1$ the differential resistivity
$r_{xx}$ at all three strong dc biases ($\omega_j/\omega =0.37, 0.56$, and 0.93)
can become negative,  implying possible resistance suppression 
or quasi-ZRSs induced by the bias current. These results indeed show up in a recent
 experimental observation.\cite{WZ-prl07}

This work was supported by the projects of the National Science Foundation of China,
the special Funds for Major State Basic Research Project, 
and the Shanghai Municipal Commission of Science and Technology.

\end{document}